Molecular insights into Neotropical bird – tick ecological associations and the role of birds in tick-borne disease ecology

Matthew J. Miller[a], Helen J. Esser[a,b], Jose R. Loaiza[a,c,d], Edward Allen Herre[a], Celestino Aguilar[a,c], Diomedes Quintero[e], Erick Alvarez[a,d], Eldredge Bermingham[a,f]

[a] Smithsonian Tropical Research Institute, Apartado Postal 0843-03092, Panamá, Republic of Panama

[b] Department of Environmental Sciences, Wageningen University, PO Box 47, 6700 AA Wageningen, the Netherlands

[c] Instituto de Investigaciones Científicas y Servicios de Alta Tecnología, Clayton, Panamá, República de Panamá

[d] Programa Centroamericano de Maestría en Entomología, Vicerrectoría de Investigación y Postgrado, Universidad de Panamá, República de Panamá

[e] Museo de Invertebrados G. B. Fairchild, Estafeta Universitaria 0824, Universidad de Panamá, República de Panamá

[f] Patricia and Phillip Frost Museum of Science, Miami, FL 33129, USA

Corresponding Author: Matthew J. Miller, millerma@si.edu

ABSTRACT

Ticks are important vectors of emerging zoonotic diseases. While adults of many tick species parasitize mammals, immature ticks are often found on wild birds. In the tropics, difficulties in species-level identification of immature ticks hinder studies of tick ecology and tick-borne disease transmission, including any potential role for birds. In Panama, we found immature ticks on 227 out of 3,498 birds representing 93 host species, about 1/8$^{th}$ of the entire Panamanian terrestrial avifauna. Tick parasitism rates did not vary with temperature or rainfall, but parasitism rates did vary with host ecological traits: non-migratory residents, forest dwelling birds, bark insectivores, terrestrial foragers and lowland species were most likely to be infested with ticks. Using a molecular library developed from adult ticks specifically for this study, we identified 130 immature ticks obtained from wild birds, corresponding to eleven tick species, indicating that a substantial portion of the Panamanian avifauna is parasitized by a variety of tick species. Furthermore, we found evidence that immature ticks show taxonomic or ecological specificity to avian hosts. Finally, our data indicate that Panamanian birds are not parasitized regularly by the tick species responsible for most known tick-borne diseases. However, they are frequent hosts of other tick species known to carry a variety of rickettsial parasites of unknown pathogenicity. Given the broad interaction between tick and avian biodiversity in the Neotropics, future work on emerging tropical tick-borne disease should explicitly consider wild birds as vertebrate hosts.

## INTRODUCTION

Wild birds are increasingly recognized as playing an important role in human and animal health. Emerging zoonotic diseases such as avian influenza, West Nile Virus, and Lyme disease (which is transmitted by ticks) often have wild birds in their transmission cycle (1-3). Wild birds can act as reservoir hosts in endemic areas (4). They also have the potential to introduce diseases into previously naïve populations by spreading pathogens and/or their vectors over large distances via migratory flyways (5,6).

Ticks (Ixodidae) are haematophagous ectoparasites that are regularly found on wild birds, providing them habitat and blood-meal resources, and are known to transmit to humans a larger variety of pathogens than any other arthropod vector group (7). Migratory birds captured in temperate regions upon their return from wintering grounds have been observed infected with larval and nymphal stages of tropical tick species (5). The ability of migratory birds to move tropical ticks over long distances, in concert with global climate change, potentially exposes extra-tropical regions to novel tropical tick-borne pathogens and vice versa (7, 21, 22). Therefore, empirical studies are needed that evaluate which tick species are frequently involved in bird parasitism and which ecological characteristics of bird species are related to increased tick infestation levels.

In the New World tropics, the greatest risk of tick-borne disease comes from *Rickettsia rickettsii*, the etiological agent of Rocky Mountain Spotted Fever. Here, its primary Neotropical vector is *Amblyomma cajennense sensu lato* (8), although *R. rickettsii* has also recently been discovered in *A. aureolatum* in Brazil (9). Recent improvements in molecular detection of tick endosymbionts have uncovered many novel *Rickettsia* species of unknown pathogenicity in Neotropical *Amblyomma* (10-12). Although adults of the genus *Amblyomma* and several other Neotropical tick species typically exploit mammals, or reptiles and amphibians to a lesser degree (13), immature forms are routinely found on birds (7, 12-19) including, rarely, nymphs of the *A. cajennense* species complex (20).

Our understanding of Neotropical bird-tick associations and hence their role in disease transmission has been hampered by species identification problems; most immature Neotropical ticks, especially those of the genus *Amblyomma*, are not readily identifiable to species by morphology alone. *Amblyomma* species diversity peaks in the Neotropics, where taxonomic keys serve to identify the nymphal stage of few *Amblyomma* species (23). For example, in a recent survey of ticks found parasitizing humans in Panama, only 38% of the recovered specimens of *Amblyomma* could be identified to species (24). Similarly, in a study documenting tick infestation patterns in wild birds from southeastern Brazil, nearly 48% of the specimens of *Amblyomma* could not be identified to species (20). This study and others (12, 14) demonstrate that new tools and approaches are needed to properly assess the role of wild birds in tick ecology and tick-borne disease transmission in the Neotropics.

Here, we: 1) identify the ecological characteristics of Panamanian bird species frequently parasitized by ticks, and determine if climate (e.g., rainfall and temperature) affects tick parasitism rates; 2) produce a robust DNA barcode library capable of identifying most of the commonly encountered ticks in Panama; 3) use the DNA barcode library to identify to species immature ticks collected off of Panamanian wild birds and 4) provide the first analysis of bird-tick ecological associations at the species level in the northern Neotropics, including an assessment of the potential role that birds may have in the transmission of tick-borne pathogens.

## RESULTS

### Ecological traits of Panamanian wild birds parasitized by ticks

We evaluated patterns of tick parasitism in 3,498 bird specimens from the STRI bird collection, representing 384 species, i.e. nearly half of the roughly 800 non-aquatic bird species recorded from Panama. Ticks parasitized a total of 227 specimens of Panamanian birds (6.5%; **Supplementary Table S1**) representing 93 avian species and 24 families. Among avian families with at least 25 specimens evaluated, the families with the greatest proportion of specimens parasitized by ticks (%) were: Thamnophilidae (antbirds: 18%; 12 of 20 species); Furnariidae (ovenbirds and woodcreepers: 15%; 11 of 20 species); Polioptilidae (gnatcatchers: 15%; 1 of 3 species); Turdidae (thrushes: 15%; 7 of 15 species); and Troglodytidae (wrens: 14%; 9 of 16 species). Resident birds (i.e. species that breed in Panama) were 3.8 times as likely to be parasitized by ticks compared to non-breeding long-distance migrant species (Fisher's exact test, $P = 0.001$). Although females had a slightly higher prevalence of tick parasitism than males, the difference was not significant (Fisher's exact test, $P = 0.23$). Among resident birds, a logistic model indicated that only forest habitation, terrestrial foraging, bark insectivory, and lowland residency were significantly associated with tick parasitism (**Table 1**).

Among 26 sites with at least 20 sampled birds, we found no relationship between the frequency of tick parasitism and annual mean temperature, temperature seasonality, annual precipitation, or precipitation seasonality, however we did recover an affect of taxonomic composition, specifically, the proportion of the sampled avifauna belonging to the five families most frequently parasitized by ticks (**Supplementary Table S2**).

### DNA identifications of adult ticks agree with morphological taxonomy

The 96 individuals in the adult reference library of morphologically identified ticks from central Panama formed 20 clusters with pairwise Kimura-2 parameter (K2P) genetic distances greater than 5% (**Supplementary Figure 1**). All 96 could be placed in clusters in agreement with the original morphological identification of the voucher with bootstrap support values of at least 99%. Among the 20 species -clusters, average nearest-neighbor K2P distance to another cluster was 15.6% and the minimum nearest-neighbor K2P distance was 12.5% (range: 12.5% – 20.5%).

Two species contained DNA barcode sub-clusters with between-cluster sequence divergence greater than 3.0%. *Haemaphysalis juxtakochi* comprised two clusters that differed by 3.0% pairwise K2P distance, with each cluster supported by 99% bootstrap support, while *Amblyomma ovale* contained one cluster of four individuals supported by 97% bootstrap and a fifth individual that varied by an average K2P distance of 3.2%. In all both cases, individuals from both clusters were collected at a shared location, suggesting that these might represent cryptic biological species. Thus, the BIN (barcode identification number) numerical taxonomy based on DNA barcode clustes recovered identified 22 unique BINs in our adult dataset; representing the 20 clusters that agree with our morphological named species as described above, as well as second BINs for both *H. juxtakochi* and *A. ovale*.

## DNA identifications of immature ticks from birds

We generated useable DNA barcode sequences from 130 immature ticks out of a total of 172 samples attempted (76% success rate). Two failures were due to double peaks in the electropheragram recovered in multiple amplification and sequencing attempts (MJM2941-T01 and MJM4264-T01); we removed these individuals from further analyses. One individual (MJM 7015) amplified its avian host (*Poecilotriccus sylvia*) DNA. The remaining 39 individuals failed in either the PCR or the sequencing step (**Supplementary Table S1**).

Sequences from the immature ticks formed 13 DNA barcode clusters. In total, 122 of 130 (94%) taxa formed 11 clusters to which we could assign a full scientific name (**Figure 2**; **Supplementary Table S1**). Thus, we can confirm that immature ticks of the following species parasitize wild birds in Panama: *H. juxtakochi*, *A. dissimile*, *A. ovale*, *A. longirostre*, *A. geayi*, *A. sabanerae*, *A. varium*, *A. calcaratum*, and *A. nodosum*. Our data establish that eight of the 18 species of *Amblyomma* ticks found in Panama parasitize birds. This includes the second global record of *A. dissimile* – a reptile and anuran specialist – parasitizing a wild bird (Mangrove Cuckoo, *Coccyzus minor*).

In total, based on the BIN alternative taxonomy, our sample of immature ticks clustered into 13 BINs, including 11 of the 22 BINs recovered in the adult reference library. Hence, we were able to assign a BIN to 100% of the samples that provided DNA sequences and to 76% (130 of 172) of all samples for which we attempted DNA barcoding. Eight other immature ticks (6%) formed two novel clusters on our phylogenetic trees; and for these we were only able to assign a BIN numerical taxonomic identification. The first BIN (ACC9360) was formed by just one immature tick that had a nearest neighbor-distance of 14.4% to the reference library cluster of *Ixodes affinis*. A second cluster (BIN: ACC9045) was formed by seven immatures whose nearest-neighbor cluster on the BOLD database did not include ticks from this study. Instead, they were most closely related to *H. leporispalustris* collected in Canada, with a nearest-neighbor distance of 6.0%. *H. leporispalustris* also occurs in Panama, but without a Panamanian adult reference sequence and given the large sequence variation, we are unable to determine whether our sample represents a genetically-divergent Panamanian *H. leporispalustris* population or distinct species of *Haemaphysalis* yet to be recorded in Panama. Thus, we refer to these two taxa as probable members of *Ixodes* and *Haemaphysalis* genera, respectively.

Species accumulation curves for our sample of immature ticks recovered from wild birds were essentially asymptotic, as were species richness estimators designed to account for unobserved species. Using the BIN taxonomy, a Chao1 species richness estimate was 13.3 (95% confidence interval: 13.0 – 19.0), compared to an observed BIN richness of 13 (**Figure 3**). A similar finding is reached using the traditional morphological taxonomy for our immature ticks: the mean Chao1 estimate for tick species richness from Panamanian birds was 11.5 (95% confidence interval: 11.0 – 19.3), compared to an observed species richness of 11 (**Supplementary Figure S2**). These results suggest that at most only a few more tick species would be recovered from wild birds in Panama given considerably greater sampling effort, and that their occurrence on wild birds would be exceedingly rare.

### Immature ticks show no species-level host specificity or ecological filtering of avian hosts

Immature tick species showed no measurable avian species host specificity. Tick species collected from at least two different bird individuals always occurred on at least two different avian host species, and tick species most frequently recovered from wild birds in our samples had the greatest number of avian host species (Pearson's *rho* = 0.984, $P < 0.0000001$). In 5 of 11 birds from which we sequenced multiple individual ticks, we found more than one species or more than one haplotype of tick, suggesting multiple different colonizations of the host.

In addition to the lack of species-level host specificity, we observed no evidence of ecological filtering in the patterns of bird-tick associations. This is visually confirmed by our quantitative species interaction network, which demonstrates broad and varied interactions among tick and bird species (**Figure 4**). We found no difference in the frequency of forest versus non-forest hosts among our 13 tick BINS ($G$= 12.369, d.f.=12, $P$=0.42), nor in the frequency of arboreal-foraging hosts ($G$= 22.18, d.f.=12, P=0.036, Bonferonni-corrected $\alpha = 0.013$). Likewise neither the proportion of bark insectivorous ($G$= 10.287, d.f.=12, P=0.59) or montane hosts ($G$= 17.996, d.f.=12, $P$=0.12) varied among tick species. We found identical results using traditional tick taxonomy.

## DISCUSSION

### Patterns of tick parasitism on Panamanian wild birds

Tick parasitism of avian hosts in Panama is relatively rare; only 6.5% of all birds examined carried ticks. Nonetheless, we observed ticks on 93 of the 384 species we examined, representing 24 avian families suggesting that parasitism by ticks includes a diverse taxonomic array of avian hosts. Marini *et al*. (14) reported for a tick parasitism rate of 17% for 313 netted birds in the Atlantic Forest of southeastern Brazil, excluding non-passeriform species except hummingbirds. Likewise, Ogrzewalska and coworkers found that 40% of 331 forest birds in Amazonian Brazil (15), and 13% of 1,725 birds from a lowland area of São Paulo (16) were infested with ticks; again these studies focused almost exclusively on passerine birds. Restricting the Panamanian dataset to include only resident passerine species found in mainland lowland forests resulted in 145 out of 1,296 Panamanian specimens parasitized (11.2%), suggesting similar rates of tick parasitism among Neotropical passerine bird communities.

We found no evidence that tick parasitism by site varied by either temperature or rainfall. However, important ecological traits appear to modulate tick parasitism rates among Panamanian birds. We found that lowland, forest inhabiting, ground and bark foraging birds were more likely than others to be infested with ticks. Prior surveys of tick parasitism on Neotropical wild birds usually did not attempt to correlate avian ecological traits with tick prevalence, but see Marini *et al*. (14). Our study was concordant with Marini and coworkers for some ecological traits correlated with tick parasitism rates, (e.g. bark insectivory), however, that study found similar infestation rates between lowland and montane birds.

### DNA barcoding of adult and immature ticks

Our findings demonstrate that DNA barcodes are a reliable method to identify Panamanian hard ticks (Ixodidae) to species, and can overcome the frequently-cited difficulties in identifying immature forms of Neotropical ticks to species using only morphological characters (15, 16, 27). Alternatives to molecular identification of immature ticks include rearing immatures to life stages that can be reliably identified to species (e.g. 20), but this is time consuming, requires special laboratory conditions that vary among species, and most immature ticks die before reaching an identifiable life stage (16, 20). DNA barcoding appears to yield a much greater percentage of successful species identifications. We were able to identify ~ 70% of immature ticks, whereas a rearing study from Brazil was able to identify only 12% of the immatures (16).

### The lack of host specificity and ecological filtering in bird-tick interactions

Although certain ecological factors are correlated with the likelihood of tick parasitism by wild birds, we found little evidence that immature ticks are actively selecting or filtering avian hosts by species, for those tick species found on birds. Apparently, tick species that infect wild birds in Panama are opportunistic parasites. This result was robust both for particular host species and for parasites among birds within ecological guilds. Previously, a study from Brazil (20) suggested that *A. longirostre* showed a preference for arboreal passerines while *A. calcaratum* and *A. nodosum* favored terrestrial passerine species, although explicit statistical tests were not employed. More importantly, our results indicate that although the adults of many Neotropical tick species show specificity for certain mammalian hosts (13, 28, 29), this does not translate to similar behavioral patterns among free-living larvae and nymphs. Thus, parameters used in tick-borne disease transmission models based on patterns of host specificity of adult ticks may require re-examination in order to incorporate the more labile ecology of immature ticks.

### The role of wild birds in the transmission ecology of tick-borne pathogens

Our ability to identify immature ticks from Panamanian birds permits insight to the role of wild birds in the transmission ecology of Neotropical tick-borne diseases. For example, our results indicate that migratory birds are relatively unimportant actors in Panamanian bird-tick interactions. Migratory birds were four times less likely to be parasitized than resident species. Interestingly, the rate of parasitism among migrant birds (< 2%) is quite similar to that found in surveys of wild birds from temperate North America (e.g. 3, 5). Furthermore, the immature ticks that we observed on migratory birds were all identified as local tick species not reported to be involved in the transmission of known human diseases (**Figure 2**). Our results suggest a relatively low risk of long distance movement of Neotropical tick-borne pathogens by migratory birds, despite evidence for occasional long distance translocation of ticks by birds in other studies (3,5,6).

We found no evidence that wild birds are involved in the transmission ecology of Rocky Mountain Spotted Fever (RMSF). RMSF is the most virulent tick-borne disease known in the Western Hemisphere and is caused by infection from *Rickettsia rickettsii*. In Panama, RMSF was first reported over 60 years ago (30), although it remained unreported again until 2004 (31), when it resulted in a fatal case in western Panama. Since 2004, RMSF has been regularly reported in central and western Panama (32). Several species of ixodid ticks are confirmed vectors of *R. rickettsii*, including three species of Panamanian ticks: *D. nitens*, *R. sanguineus* and *A. mixtum*. However, members of the *A. cajennense* species complex (in Panama: *A. mixtum*) are considered the primary vector of *R. rickettsii* in tropical America (8), and in Panama *R. rickettsii* has only been detected in *A. mixtum* (33, 34). We found no examples of *A. mixtum*, *D. nitens*, or *R. sanguineus* parasitizing birds in our study. Furthermore, our species accumulation curves (**Figure 3, Supplementary Figure S2**) for tick species found on wild birds suggest that at best only a few, rare, species are yet to be recovered from resident wild land birds in Panama. A few studies from other Neotropical regions have reported that immature forms of species in the *A. cajennense* complex parasitize various wild bird species, however Labruna et al. (20) challenged the morphological identifications in several cases. That study and others (15, 16, 18), including ours, demonstrate that *A. mixtum*, *D. nitens*, and *R. sanguineus* are at best extremely rare parasites of wild birds, and collectively suggest a negligible role for wild birds in Rocky Mountain Spotted Fever transmission in Panama and elsewhere in the Neotropics.

Although we did not attempt to isolate *Rickettsia* from sampled ticks, other studies have demonstratd that all but one of the tick species found to parasite Panamanian wild birds harbor *Rickettsia*. *Rickettsia amblyommii* has been suggested, but not confirmed, to be a cause of spotted fever-like disease in North America (35) and has been recovered as a parasite of *A. ovale* in Panama (36, 37) as well as *A. geayi* and *A. longirostre* in Brazil (11, 15). *Rickettsia parkeri*, which was recently identified as the cause of human spotted fever rickettsiosis in southeastern USA as well as Brazil, Uruguay and Argentina was recovered from *A. nodosum* and *A. ovale* in Brazil (11). Furthermore, other *Rickettsia* species, not yet known to cause human disease, have been recovered from *A. calcaratum* (38), *A. dissimile* (39), *A. varium* (40) and *H. juxtakochi* (11). Work in the Neotropical regions concerning the pathogenicity of *R. amblyommii* and *R.*

*parkeri* and other rickettsiales is in its infancy. Likewise, spotted fever group rickettsioses are often under-detected, especially in Middle and South America (40). Thus, our findings of a pervasive and diverse relationship between birds and immature ticks in Panama suggests that further consideration of the role of wild birds in the ecology of tick-borne disease is warranted.

## MATERIALS AND METHODS

### Ecological patterns of tick parasitism on birds

We generated a tick parasitism data matrix (presence or absence of hard ticks: Ixodidae) from the ectoparasite collection of bird specimens collected between 2008 and 2013 throughout Panama (**Figure 1**) maintained by the Smithsonian Tropical Research Institute Bird Collection (STRIBC). Non-land birds from marsh, aquatic, and riverine habitats (e.g. Anseriformes, Charadriiformes, Alcedinidae) were excluded from the analysis, as well as aerial foragers (e.g. Apodidae, Hirundinidae), resulting in 3498 specimen records (**Supplementary Table 1**). Per STRIBC field protocols, wild birds were euthanized in the field and flash frozen on solid $CO_2$ in individual freezer bags to eliminate the risk of cross-contamination of ectoparasites prior to transportation to the lab. In the lab, during the initial stages of specimen preparation, ectoparasites were removed from the frozen bird carcass and transferred into vials containing 95% ethanol. We made no attempt to identify bird-collected immature ticks prior to molecular analysis.

Using this matrix, we analyzed whether key traits of the host species were correlated with tick parasitism using contingency table analysis. We considered the relationship of eight traits on tick parasitism, some of which have previously been correlated with tick parasitism in birds: residents vs. non-breeding migrants, females vs. males, terrestrial foraging vs. arboreal, ground cavity nesting, tree hole nesting, lowland vs. montane habitats, bark insectivory, forest vs. non-forest habitats. We evaluated the influence of sex and migratory status on the entire dataset, however because ecological traits for many migratory birds are more labile away from their breeding grounds and classifications based on temperate zone ecology may not reflect behavior in Panama, we evaluated the relationship between the remaining six traits and tick parasitism only on resident birds (i.e. those species that breed in Panama). Montane species were defined as those found almost exclusively 600 meters above sea level. Terrestrial foraging species include those species that forage primarily, but not exclusively, on the ground or within 15 centimeters of the ground. Forest inhabitants live in or require mature tropical forests; most edge species were classified as non-forest inhabitants. We evaluated all possible combinations of ecological characters using logistic regression models.

### Adult DNA Barcode Reference Library

We generated a DNA barcode reference library for the ticks of central Panama using morphologically pre-identified adult ticks collected as part of on-going research programs (HJE and JRL) on the tick-host interactions in the area surrounding the Panama Canal, and also in a few cases from ethanol preserved museum specimens of adult ticks (**Figure 1).** Field collections were accomplished either by removal of ticks from hosts (live-captured animals, road kill, livestock, and pets) or by collecting questing ticks from the free environment via flagging, e.g. sweeping a white cotton cloth along vegetation and leaf litter and harvesting accumulated ticks. Adults were identified using morphological characters and existing taxonomic keys (13, 41), and were stored in 95% ethanol and frozen at -20ºC prior to molecular analysis (see next section). Our adult reference library included 96 individuals (**Supplementary Table 2**) that were morphologically assigned to 19 of the approximately 37 species of hard ticks recognized for the Republic of Panama, including 14 of 18 species of *Amblyomma* (13). Unless already part of a museum collection, after DNA extraction adult reference ticks were stored in 95% ethanol and are maintained as voucher specimens in the ectoparasite collection of the STRIBC. A public dataset for these 96 specimens, including geographic details of collection, specimen photographs, and museum voucher information can be found on the BOLD data portal v3 (42) under the name: DS-TICKA (dx.doi.org/10.5883/DS-TICKA).

### Immature Ticks from Panamanian Wild Birds

We selected 186 immature ticks from the pool of immature ticks collected from birds for molecular species-level identification using the adult reference DNA barcode library as the basis for identification (**Supplementary Table 1**). We obtained usable DNA barcode sequences for 130 (see Results). Sample details for immature ticks can be found on the BOLD data portal under the name: DS-TICKI (dx.doi.org/10.5883/DS-TICKI).

## Molecular methods

To allow for the preservation of museum vouchers, DNA was extracted from adult specimens from either two legs removed from the specimen, or from a rear quarter section of the abdomen cut from the body, done under an entomological dissecting microscope. We used the entire body of immature ticks for DNA extraction. In all cases, the material being extracted was frozen in a 2 ml tube suspended in liquid nitrogen and pulverized using a sterile micro-pestle to improve DNA yield. We initially obtained poor DNA yield after attempted DNA extractions using DNAeasy spin columns (Qiagen, Valencia, CA), following the manufacturer's instructions (except that we reduced the final elution volume to 50 µl). Subsequently, we switched to the QIAamp DNA Micro kit (Qiagen), which uses similar spin column technology but is optimized for smaller samples and resulted in superior DNA yields.

Amplification of the DNA barcoding region (5' region of the COI mitochondrial gene, 43) was accomplished using the standard invertebrate primers (LCO1490 and HCO2198; 44) following (45), except that we halved the reaction volume (i.e. 25 µL) and raised DNA to 4 µL; we used Qiagen taq and buffers. Positive and negative controls were run in every reaction. Amplifications were visualized on a low-melting agarose gel from which a single PCR product was extracted using a sterilized scalpel blade, and sequenced at the Naos Molecular Laboratory, Smithsonian Tropical Research Institute. DNA sequences and tracefiles can be examined in BOLD under the DS-TICKA database and DS-TICKI database. Sequences have also been deposited in GenBank under accession numbers KF200076 – KF200171.

## Tree building and barcode distance analysis

In order to understand species limits and confirm morphological identification among our adult reference ticks (*N* = 96), we generated a neighbor-joining tree in MEGA v.5.1 (46) using Kimura-2 parameter (K2P) distances. Branch support was assessed by bootstrapping the topology with 500 replicates. We examined the K2P distance matrix and resulting topology for evidence of genetic divergence among our adult reference library that might provide evidence for the presence of cryptic species (e.g. 47) using both a standard genetic distance approach (3% K2P) as well as looking for the assignation of multiple Barcode Index Numbers (BINs) to a given species. The Barcode Index Number is an alternative, numerical taxonomy that clusters taxa into interim operational taxonomic units using a stage process to employ single linkage clustering (48). BINs are assigned automatically in the BOLD database portal based on the global dataset of DNA barcode sequences (i.e. including samples not generated in this study). We repeated all tree-building, genetic distance, and clustering analyses for a second, expanded dataset that combined the 96 adult reference sequences with 130 sequences from immature ticks collected from birds.

### Species-level associations between ticks and wild birds in Panama

We used the identifications of immature ticks from STRIBC bird specimens to further examine species-specific bird-tick associations in Panama. First, we assessed host-specificity by examining the correlation of the number of tick individuals recovered and the number of avian host species for all tick species recovered in our dataset; if immature ticks are non-host specific, this correlation should be strong. While this approach provides one measurement of host–tick specificity, it potentially overlooks the role of ecological filtering of hosts for particular parasite species (e.g. 49, 50), which we tested for by examining differences in the frequency of parasitism by host ecological traits for each tick species identified using COI barcodes using *G*-tests of independence. We visualized immature tick species-specific associations with ecological traits of avian hosts via quantitative interaction networks created using the bipartite package following the authors instructions (51) in the R statistical application (52). For those birds that had multiple ticks identified by DNA barcoding we scored each tick species separately but only once in the data matrix.

Finally, to estimate the proportion and distribution of the total Panamanian tick species pool that might depend on birds as host vertebrates for immature life stages, we estimated the total species richness of ticks that parasitize wild birds in Panama through species accumulation curves generated in the EstimateS software package (53). When sampling is exhaustive, the species accumulation curve should reach an asymptote. However, even non-exhaustive sampling can still yield sufficient data to provide a reasonable estimate of the true species richness, which can be assessed by observing an asymptote in the statistical estimate of species richness (54). We generated species-accumulation curves (SACs) for the 130 ticks where we were successfully able to generate DNA barcodes using both the adult reference library cluster-indicated taxonomy, and using the BIN numerical taxonomy generated in the BOLD database. In both cases, we used the Chao1 species richness estimator (55), which attempts to non-parametrically correct the observed species richness as a function of the proportion of species observed exactly once or twice in the dataset. Mean Chao1 values were obtained from 100 reshuffles of our data set with replacement in EstimateS; sampling with replacement being critical in order to account for sampling error.

## ACKNOWLEDGEMENTS

We thank Panama's Environmental Agency (ANAM) for granting us bird and tick collecting permits, without which this study would not have been possible. We thank the members of the Smithsonian Tropical Research Bird Collection field team, as well as B. Voirin and B. Hirsch for providing adult ticks used in this study, and P. Jansen for comments. Funding was provided by an inter-agency award from the US Centers for Disease Control ("Effect of Anthropogenic Climate Change on the Ecology of Zoonotic and Vector-borne Diseases"), a Smithsonian Grand Challenges Award ("Tropical Vertebrate Diversity Loss and the Emergence of Tick-borne Diseases." HLE was supported by the graduate school of Production Ecology and Resource Conservation of Wageningen University. The STRI Bird Collection was supported by an NIH/NSF Ecology and Evolution of Infectious Diseases Award from the Fogarty International Center (3R01-TW005869-05S1). DNA barcode sequencing was supported from a grant from the Smithsonian Institution's DNA Barcoding Network.

## SUPPLEMENTARY MATERIAL

**Supplementary Table 1.** Host species, collecting location, number of ticks recovered, and results of tick DNA barcoding (when applicable) for 227 Panamanian birds parasitized by ticks. Available at: http://dx.doi.org/10.6084/m9.figshare.1250139.

**Supplementary Table 2.** Specimen number, collection data, and morphological identification for 96 adult ticks used in reference library.

## REFERENCES

1. Rappole JH, Hubálek Z (2003) Migratory birds and West Nile virus. *J Appl Microbiol* 94(s1):47-58.
2. Kilpatrick AM, et al. (2006) Predicting the global spread of H5N1 avian influenza. *Proc Natl Acad Sci U S A* 103(51):19368-19373.
3. Ogden NH, et al. (2008) Role of migratory birds in introduction and range expansion of *Ixodes scapularis* ticks and of *Borrelia burgdorferi* and *Anaplasma phagocytophilum* in Canada. *Appl Environ Microbiol* 74(6):1780-1790.
4. Ginsberg HS, et al. (2005). Reservoir competence of native North American birds for the Lyme disease spirochete, *Borrelia burgdorferi*. *J Med Entomol* 42(3): 445-449.
5. Hamer SA, et al. (2012) Wild birds and urban ecology of ticks and tick-borne pathogens, Chicago, Illinois, USA, 2005–2010. *Emerg Infect Dis* 18(10):1589.
6. Lindeborg M, et al. (2012) Migratory birds, ticks, and Crimean-Congo hemorrhagic fever virus. *Emerg Infect Dis* 18(12):2095.
7. Jongejan F, Uilenberg G (2004) The global importance of ticks. *Parasitology* 129:S3-S14.
8. Dantas-Torres F (2007) Rocky Mountain spotted fever. *Lancet Infect Dis* 7:724 - 732.
9. Pinter A, Labruna MB (2006) Isolation of *Rickettsia rickettsii* and *Rickettsia bellii* in cell culture from the tick *Amblyomma aureolatum* in Brazil. *Ann N Y Acad Sci* 1078:523-529.
10. Labruna MB, et al. (2004) Molecular evidence for a spotted fever group *Rickettsia* species in the tick *Amblyomma longirostre* in Brazil. *J Med Entomol* 41:533-537.
11. Labruna MB et al. (2011) Rickettsioses in Latin America, Caribbean, Spain and Portugal. *Revista MVZ Córdoba* 16:2435-2457.
12. Ogrzewalska M, Martins T, Capek M, Literak I, Labruna MB (2013) A *Rickettsia parkeri*-like agent infecting *Amblyomma calcaratum* nymphs from wild birds in Mato Grosso do Sul, Brazil. *Ticks Tick Borne Dis* 4:145-7.
13. Fairchild GB, Kohls GM, Tipton VJ (1966) in *Ectoparasites of Panama*, eds Wenzel RL, Tipton VJ (Field Museum of Natural History, Chicago), pp 167-219.
14. Marini MA, Reinert BL, Bornschein MR, Pinto JC, Pichorim MA (1996) Ecological correlates of ectoparasitism on Atlantic Forest birds, Brazil. *Ararajuba* 4:93-102.

15. Ogrzewalska M, Uezu A, Labruna MB (2010) Ticks (Acari: Ixodidae) infesting wild birds in the eastern Amazon, northern Brazil, with notes on rickettsial infection in ticks. *Parasitol Res* 106:809-816.

16. Ogrzewalska M, et al. (2009) Ticks (Acari: Ixodidae) infesting birds in an Atlantic rain forest region of Brazil. *J Med Entomol* 46:1225-1229.

17. Ogrzewalska M, Uezu A, Jenkins CN, Labruna MB (2011) Effect of forest fragmentation on tick infestations of birds and tick infection rates by rickettsia in the Atlantic forest of Brazil. *Ecohealth* 8:320-31.

18. Pacheco RC, et al. (2012) Rickettsial infection in ticks (Acari: Ixodidae) collected on birds in southern Brazil. *J Med Entomol* 49:710-716.

19. Ogrzewalska M, Literak I, Cardenas-Callirgos JM, Capek M, Labruna MB (2012) *Rickettsia bellii* in ticks *Amblyomma varium* Koch, 1844, from birds in Peru. *Ticks Tick Borne Dis* 3:254-6.

20. Labruna MB, et al. (2007) Ticks collected on birds in the state of São Paulo, Brazil. *Exp Appl Acarol* 43:147-60.

21. Parola P, et al. (2008) Warmer weather linked to tick attack and emergence of severe rickettsioses. *PLoS Negl Trop Dis* 2:e338.

22. Randolph SE (2010) To what extent has climate change contributed to the recent epidemiology of tick-borne diseases? *Vet Parasitol* 167:92-4.

23. Martins TF, Onofrio VC, Barros-Battesti DM, Labruna MB (2010) Nymphs of the genus *Amblyomma* (Acari: Ixodidae) of Brazil: descriptions, redescriptions, and identification key. *Ticks Tick Borne Dis* 1:75-99.

24. Bermúdez C SE, et al. (2012) Ticks (Ixodida) on humans from central Panama, Panama (2010-2011). *Exp Appl Acarol* 58:81-8.

25. Keesing F, et al. (2010) Impacts of biodiversity on the emergence and transmission of infectious diseases. *Nature* 468:647-52.

26. Schmidt KA, Ostfeld RS (2001) Biodiversity and the dilution effect in disease ecology. *Ecology* 82:609-619.

27. Bermúdez SE, Miranda RJ, Smith D (2010) Ticks species (Ixodida) in the Summit Municipal Park and adjacent areas, Panama City, Panama. *Exp Appl Acarol* 52:439-48.

28. Guglielmone AA, et al. (2003) *Amblyomma aureolatum* (Pallas, 1772) and *Amblyomma*

*ovale* Koch, 1844 (Acari: Ixodidae): hosts, distribution and 16S rDNA sequences. *Vet Parasitol* 113:273-288.

29. Nava S, Szabó MPJ, Mangold AJ, Guglielmone AA (2008) Distribution, hosts, 16S rDNA sequences and phylogenetic position of the Neotropical tick *Amblyomma parvum* (Acari: Ixodidae). *Ann Trop Med Parasitol* 102:409-425.

30. de Rodaniche EC, Rodaniche A (1950) Spotted fever in Panama; isolation of the etiologic agent from a fatal case. *Am J Trop Med Hyg* 30:511-7.

31. Estripeaut D et al. (2007) Rocky mountain spotted fever, Panama. *Emerg Infect Dis* 13:1763.

32. Bermúdez CS et al. (2011) Rickettsial infection in domestic mammals and their ectoparasites in El Valle de Antón, Coclé, Panamá. *Vet Parasitol* 177:134-8.

33. de Rodaniche EC (1953) Natural infection of the tick, *Amblyomma cajennense*, with Rickettsia rickettsii in Panama. *Am J Trop Med Hyg* 2:696-699.

34. Estrada-Peña A, Guglielmone AA, Mangold AJ (2004) The distribution and ecological 'preferences' of the tick *Amblyomma cajennense* (Acari: Ixodidae), an ectoparasite of humans and other mammals in the Americas. *Ann Trop Med Parasitol* 98:283-292.

35. Apperson CS, et al. (2008) Tick-borne diseases in North Carolina: is "*Rickettsia amblyommii*" a possible cause of rickettsiosis reported as Rocky Mountain spotted fever? *Vector Borne Zoonotic Dis* 8(5):597-606.

36. Bermúdez SE, et al. (2009) Detection and identification of rickettsial agents in ticks from domestic mammals in Eastern Panama. *J Med Entomol* 46:856-861.

37. Eremeeva ME, et al. (2009) Spotted fever rickettsiae, *Ehrlichia* and *Anaplasma*, in ticks from peridomestic environments in Panama. *Clin Microbiol Infect* 15 Suppl 2:12-4.

38. Ogrzewalska M, Martins T, Capek M, Literak I, Labruna MB (2013) A *Rickettsia parkeri*-like agent infecting *Amblyomma calcaratum* nymphs from wild birds in Mato Grosso do Sul, Brazil. *Ticks Tick Borne Dis* 4(1):145-147.

39. Miranda J, Portillo A, Oteo JA, Mattar S (2012) *Rickettsia* sp. strain Colombianensi (Rickettsiales: Rickettsiaceae): a new proposed *Rickettsia* detected in *Amblyomma dissimile* (Acari: Ixodidae) from iguanas and free-living larvae ticks from vegetation. *J Med Entomol* 49(4):960-965.

40. Romer Y, et al. (2011) *Rickettsia parkeri* Rickettsiosis, Argentina. *Emerg Infect Dis*

17:1169-73.

41. Onofrio V, Labruna MB, Pinter A, Giacomin F, Barros-Battesti DM (2006) Comentários e chaves as espécies do gênero *Amblyomma*. *Carrapatos de Importância Médico-Veterinária da Região Neotropical: Un guia ilustrado para identificação de espécies*, eds Barros-Battesti DM, Arzua M, Bechara GH (Instituto Butantan, Sao Paulo), pp 53-113.

42. Ratnasingham S, Hebert PD (2007) Bold: The Barcode of Life Data System (http://www.barcodinglife.org). *Mol Ecol Notes* 7:355-364.

43. Hebert PD, Ratnasingham S, de Waard JR (2003) Barcoding animal life: cytochrome c oxidase subunit 1 divergences among closely related species. *Proc Biol Sci* 270:S96-S99.

44. Folmer O, Black M, Hoeh W, Lutz R, Vrijenhoek R (1994) DNA primers for amplification of mitochondrial cytochrome c oxidase subunit I from diverse metazoan invertebrates. *Mol Mar Biol Biotechnol* 3:294-9.

45. Kumar NP, Rajavel AR, Natarajan R, Jambulingam P (2007) DNA barcodes can distinguish species of Indian mosquitoes (Diptera: Culicidae). *J Med Entomol* 44:1-7.

46. Tamura K, et al. (2011) MEGA5: Molecular Evolutionary Genetics Analysis Using Maximum Likelihood, Evolutionary Distance, and Maximum Parsimony Methods. *Mol Biol Evol* 28:2731-2739.

47. Hebert PD, Penton EH, Burns JM, Janzen DH, Hallwachs W (2004) Ten species in one: DNA barcoding reveals cryptic species in the Neotropical skipper butterfly *Astraptes fulgerator*. *Proc Natl Acad Sci U S A* 101:14812-14817.

48. Ratnasingham S, Hebert PD (2013) A DNA-based registry for all animal species: the barcode index number (BIN) system. *PLoS One* 8:e66213.

49. Combes C (2001) *Parasitism: the ecology and evolution of intimate interactions*, (University of Chicago Press, Chicago).

50. Gager AB, Del Rosario Loaiza J, Dearborn DC, Bermingham E (2008) Do mosquitoes filter the access of *Plasmodium* cytochrome b lineages to an avian host? *Mol Ecol* 17:2552-61.

51. Dormann CF, Fründ J, Blüthgen N, Gruber B (2009) Indices, graphs, and null models: analyzing bipartite ecological networks. *Open Ecology Journal* 2:7-24.

52. R Development Team (2008) R: A language and environment for statistical computing. R Foundation for Statistical Computing Vienna, Austria. SBN 3-900051-07-0, URL:

http://www.R-project.org.
53. Colwell RK (2013) EstimateS: Statistical estimation of species richness and shared species from samples. Version 9. User's Guide and application published at: http://purl.oclc.org/estimates.
54. Colwell RK, Coddington JA (1994) Estimating terrestrial biodiversity through extrapolation. *Philos Trans R Soc Lond B Biol Sci* 345:101-118.
55. Chao A (1987) Estimating the population size for capture-recapture data with unequal catchability. *Biometrics* 43:783-791.

FIGURE LEGENDS

**Figure 1. Map of collecting areas for adult and immature ticks.** Blue circles represent areas where adult ticks were sampled for the reference library; orange circles represent locations where immature ticks were sampled from birds.

**Figure 2. DNA barcoding neighbor-joining tree of immature ticks and adult ticks from reference library.** Identification corresponding to traditional morphological taxonomy in blue blue (two clusters comprised of only immatures for which no adult voucher exists – marked with asterisks), whereas taxonomy following the alternative BIN numerical marked in gray. The distribution of Panamanian ticks recovered from wild birds is unbalanced towards certain species of *Amblyomma*. Importantly, we recovered no immature ticks on birds from species known to vector human disease (6).

**Figure 3a-d. Bird – immature tick quantitative interaction networks**. a) Green: interactions involving forest inhabiting bird species, yellow: interactions involving non-forest bird species; b) blue: interactions involving arboreal bird species, yellow: terrestrial-foraging bird species; c) brown: interactions involving bird species that are bark insectivores; yellow: other species; d) red: interactions involving montane bird species, yellow: lowland species. a-d) gray: interactions involving non-breeding migrant bird species. The frequency of hosts among these four ecological traits was not significant among the 11 sampled tick species.

**Figure 4. Species accumulation curve (SAC) for immature ticks recovered from Panamanian wild birds based on BIN numerical taxonomy** (see Supplementary Figure S2 for SAC based on traditional tick taxonomy). Red line = $S$, mean observed species richness; dark blue line = $\hat{S}$, mean Chao1 $S$ estimate; filled blue area defines 95% upper and lower confidence limits (CI) for $\hat{S}$. As Chao1 is downward biased, the 95% lower CI is probably not useful. The convergence of the S and $\hat{S}$ curves, as well as the asymptotic nature of these curves and the CI curves, suggests that at most only a few more tick species would be recovered from wild birds in Panama, and that their occurrence would be rare.

**Supplementary Figure S1. Phylogenetic tree of adult reference library**. Neighbor-joining tree of 96 adult ticks based on COI DNA barcode codes. Available at: http://dx.doi.org/10.6084/m9.figshare.1250141.

**Supplementary Figure S2. Species accumulation curve (SAC) for immature ticks recovered from Panamanian wild birds based BIN numerical taxonomy**. Black line = $S$, mean observed species richness; solid gray line = $\hat{S}$, mean Chao1 $S$ estimate; dotted gray lines = 95% upper and lower confidence limits (CI) for $\hat{S}$. As Chao1 is downward biased, the 95% lower CI is probably not useful. Fairchild et al (22) estimated that 37 species of hard ticks occur in Panama. Available at: http://dx.doi.org/10.6084/m9.figshare.1250142.

**Table 1. Ecological traits associated with tick parasitism among Panamanian resident wild birds (*N* = 3274).** Traits significantly associated with tick parasitism marked with (*). Significance determined by multiple logistic regression.

| Ecological Trait | Odds Ratio | *P* value (Wald's) |
|---|---|---|
| Bark insectivores* | 8.3 | 0.004 |
| Terrestrial foragers* | 4.0 | $< 10^{-15}$ |
| Forest vs. non-forest dwellers* | 3.3 | $< 10^{-9}$ |
| Tree hole nesters | 2.5 | 0.20 |
| Ground cavity nesters | 2.0 | 0.14 |
| Lowland vs. montane* | 1.7 | 0.01 |

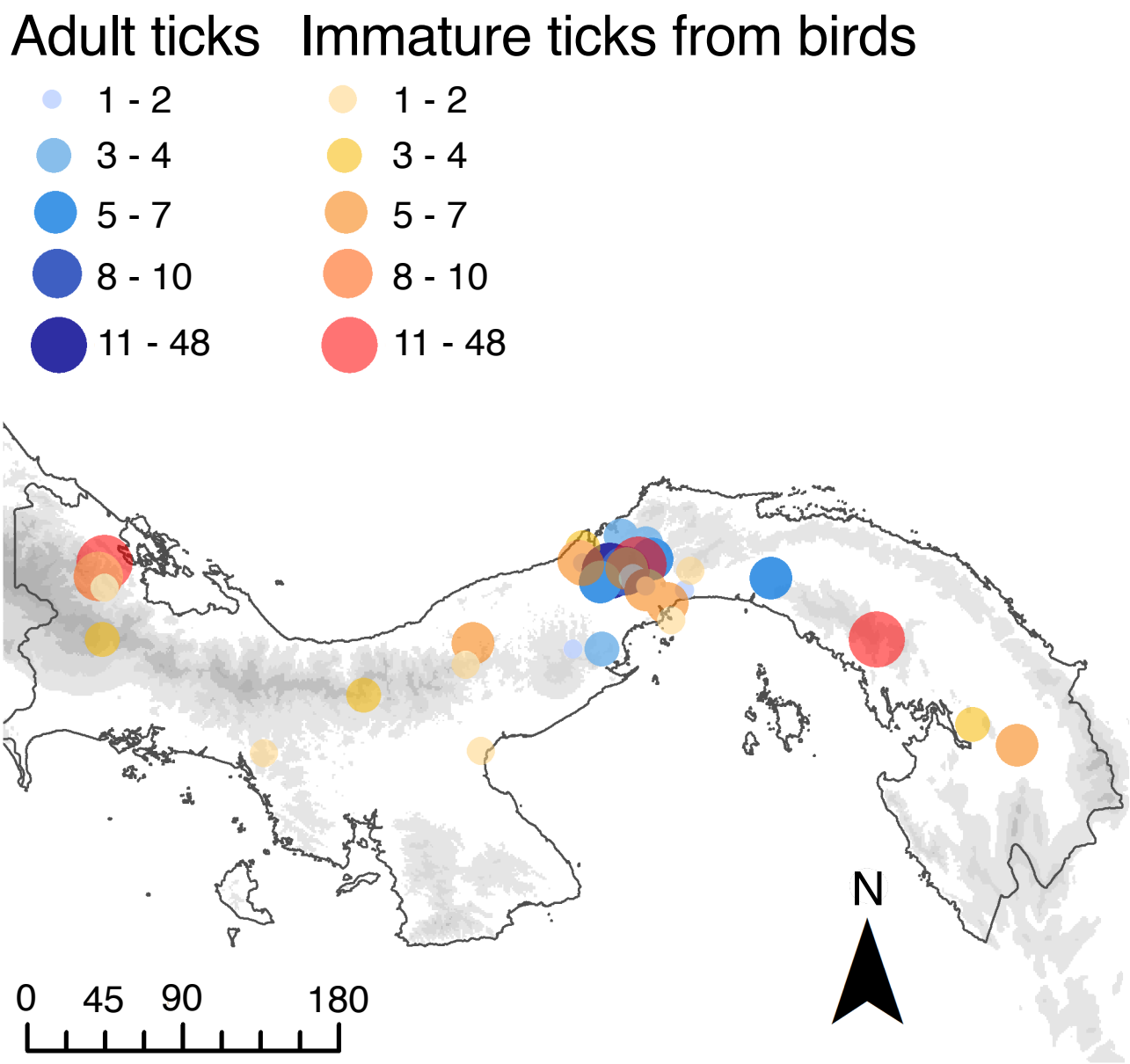
Adult ticks
1 - 2
3 - 4
5 - 7
8 - 10
11 - 48
Immature ticks from birds
1 - 2
3 - 4
5 - 7
8 - 10
11 - 48
N
0
45
90
180

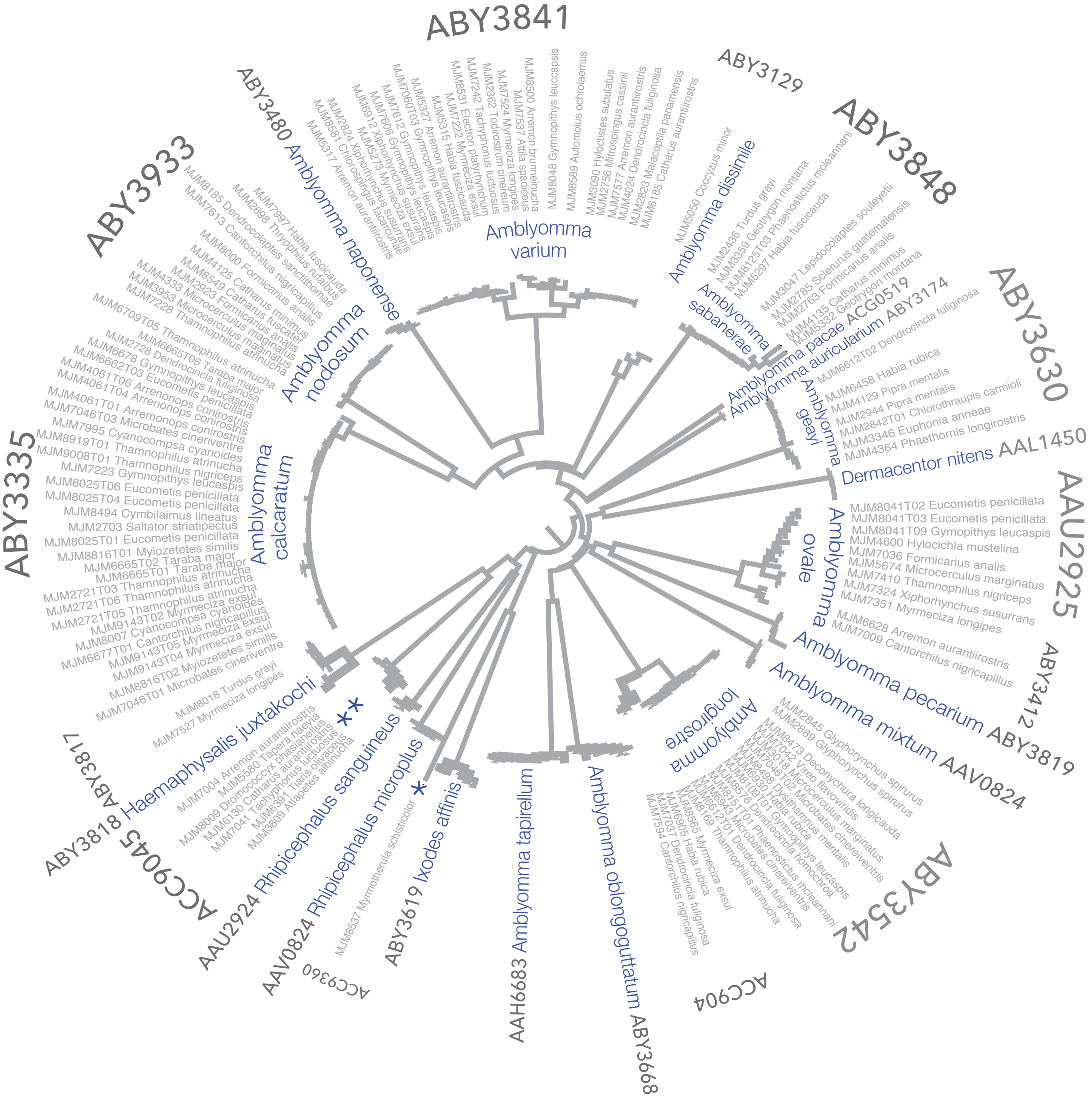

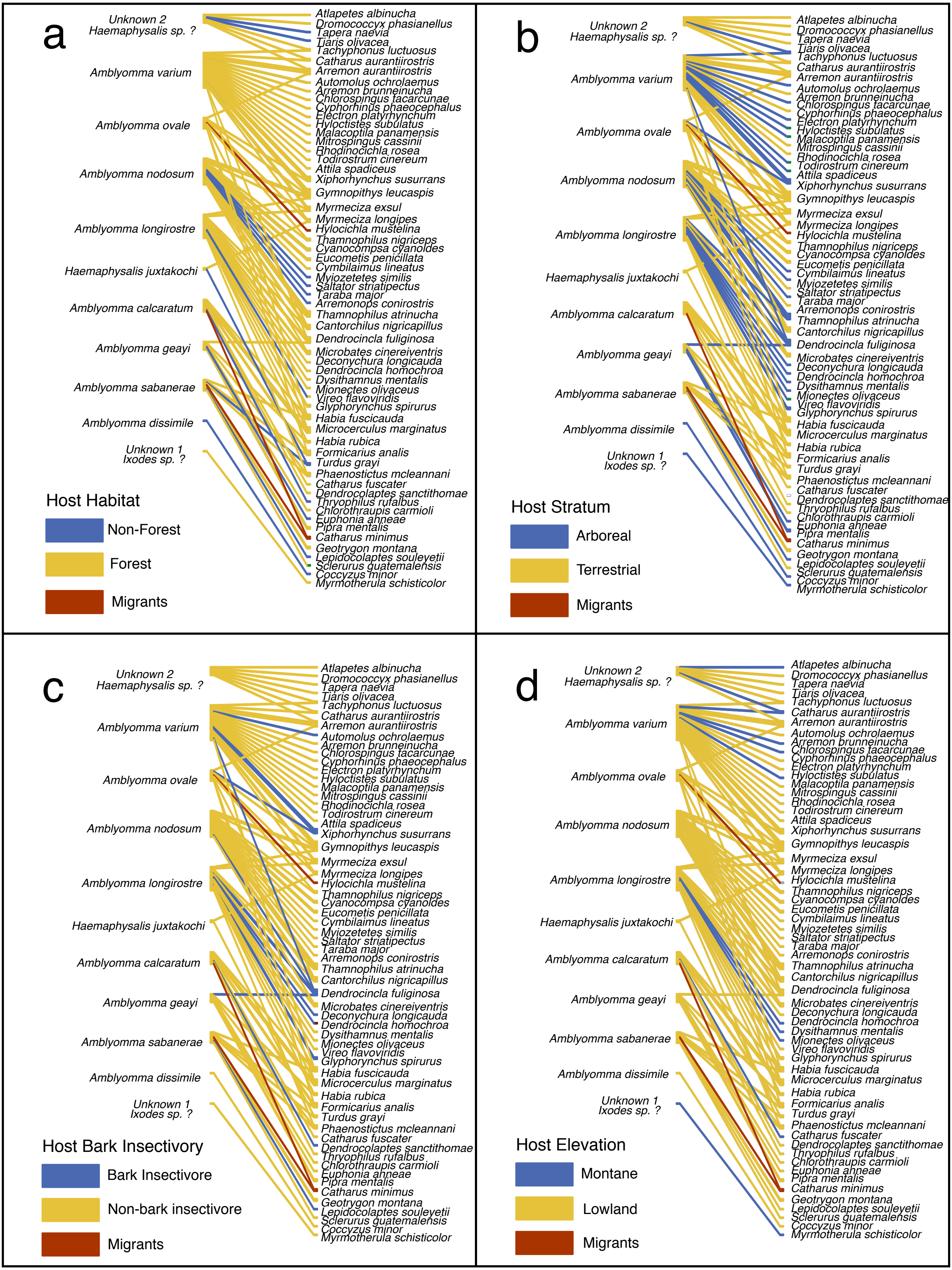

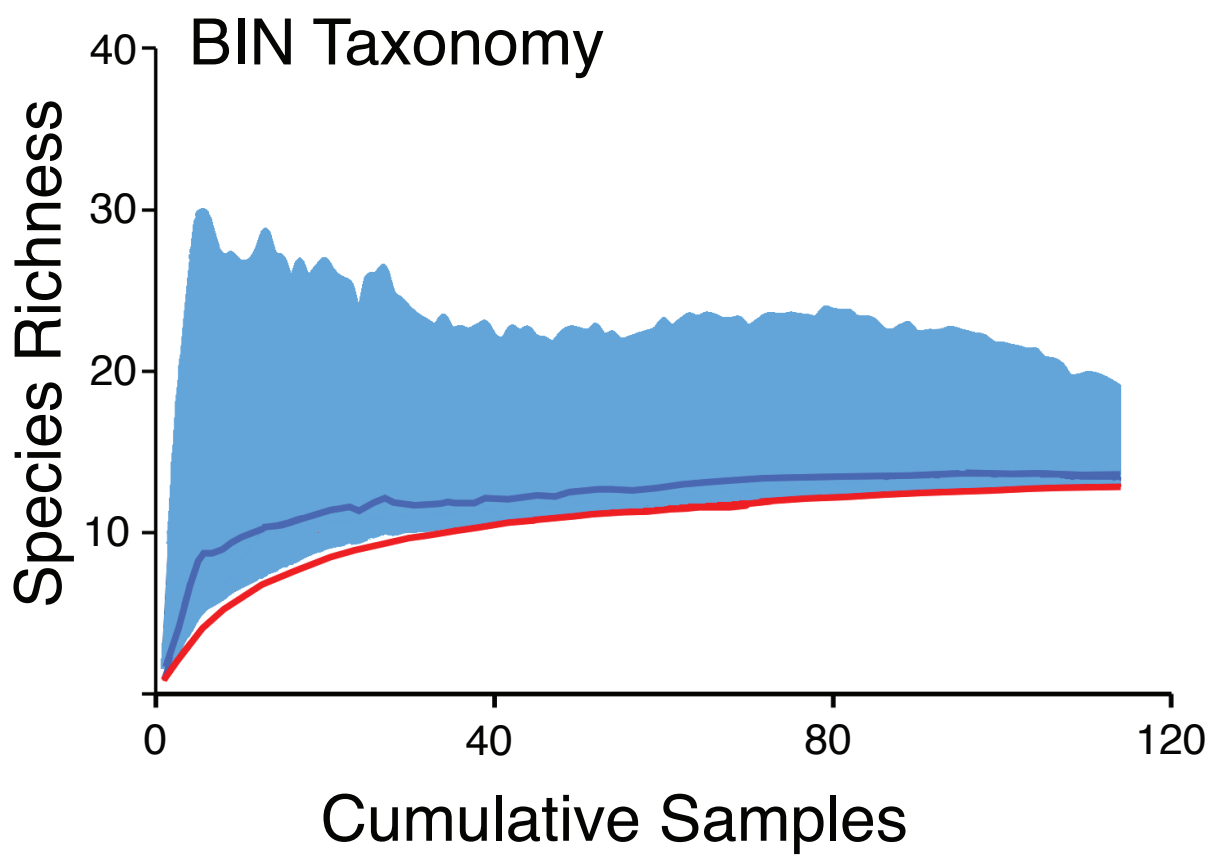